\documentclass[aastex,showpacs,superscriptaddress,showpacs,letterpaper,showkeys,preprintnumbers,altaffilletter,amssymb,amsmath,amsfonts,aps]{emulateapj}

\usepackage{aas_makros_pk}
\usepackage{textcomp}

\usepackage{natbib}
\def\xp{{\textsc{X-Pipeline}}}

\newcommand{\sci}[2]{#1 \times 10^{#2}}
\newcommand{\egw}{E_{\mathrm{GW}}}
\newcommand{\flare}{\textsc{Flare}}

\newcommand{\onsourceFAP}{76}
\newcommand\BNSDninety{2.1}
\newcommand\BNSDninetyFaceOn{5.2}
\newcommand\BNSExc{71}
\newcommand\BNSExcFaceOn{98}
\newcommand\NSBHDninety{5.3}
\newcommand\NSBHDninetyFaceOn{10.7}
\newcommand\NSBHExc{93}
\newcommand\NSBHExcFaceOn{>99}

\newcommand{\dccversion}{12}

\slugcomment{{LIGO-P1000097}-v\dccversion}

\shorttitle{Implications for the Origin of GRB~051103 from LIGO Observations}
\shortauthors{Abbott et al.}

\begin{document}
\title{Implications for the Origin of GRB~051103 from LIGO Observations}

\author{J.~Abadie$^{1}$,	
B.~P.~Abbott$^{1}$,	
T.~D.~Abbott$^{60}$,    	
R.~Abbott$^{1}$,	
M.~Abernathy$^{2}$,	
C.~Adams$^{3}$,	
R.~Adhikari$^{1}$,	
C.~Affeldt$^{4,11}$,	
P.~Ajith$^{1}$,	
B.~Allen$^{4,5,11}$
% ...
}	
\address{$^{1}$LIGO - California Institute of Technology, Pasadena, CA  91125, USA }
\address{$^{2}$University of Glasgow, Glasgow, G12 8QQ, United Kingdom }
\address{$^{3}$LIGO - Livingston Observatory, Livingston, LA  70754, USA }
\address{$^{4}$Albert-Einstein-Institut, Max-Planck-Institut f\"ur Gravitationsphysik, D-30167 Hannover, Germany}
\address{$^{5}$University of Wisconsin--Milwaukee, Milwaukee, WI  53201, USA }
% ...
\address{$^{11}$Leibniz Universit\"at Hannover, D-30167 Hannover, Germany }
% ...
\address{$^{60}$California State University Fullerton, Fullerton CA 92831, USA}
% ...

\author{M.~A.~Bizouard$^{1}$}
\author{A.~Dietz$^{2}$}
\author{G.~M.~Guidi$^{3ab}$}
\author{M.~Was$^{1}$}
\address{$^{1}$LAL, Universit\'e Paris-Sud, IN2P3/CNRS, F-91898 Orsay, France}
\address{$^{2}$Laboratoire d'Annecy-le-Vieux de Physique des Particules (LAPP),
         Universit\'e de Savoie,CNRS/IN2P3, F-74941 Annecy-Le-Vieux, France}
\address{$^{3}$INFN, Sezione di Firenze, I-50019 Sesto Fiorentino$^a$;
         Universit\`a degli Studi di Urbino 'Carlo Bo', I-61029 Urbino$^b$,
         Italy}

\date{\today}

\begin{abstract}
We present the results of a LIGO search for gravitational waves (GWs) associated
with GRB~051103, a short-duration hard-spectrum gamma-ray burst whose
electromagnetically determined sky position is coincident with the spiral galaxy M81,
which is 3.6\,Mpc from Earth. Possible progenitors for short-hard GRBs include
compact object mergers and soft gamma repeater (SGR) giant flares.  A merger
progenitor would produce a characteristic GW signal that should be detectable at
the distance of M81, while GW emission from an SGR is not expected to be
detectable at that distance.  We found no evidence of a GW signal associated
with GRB~051103. Assuming weakly beamed $\gamma$-ray emission with a jet
semi-angle of $30^{\circ}$ we exclude a binary neutron star merger in M81 as
the progenitor with a confidence of $\BNSExcFaceOn\%$. Neutron star-black hole
mergers are excluded with $\NSBHExcFaceOn\%$ confidence. If the event occurred
in M81 our findings support the hypothesis that GRB 051103 was due to an SGR
giant flare, making it the most distant extragalactic magnetar observed to date.
\end{abstract}




\section{Introduction} 
\label{sec:intro} 

\begin{figure}[!t]
\begin{center}
\includegraphics[angle=0,width=85mm]{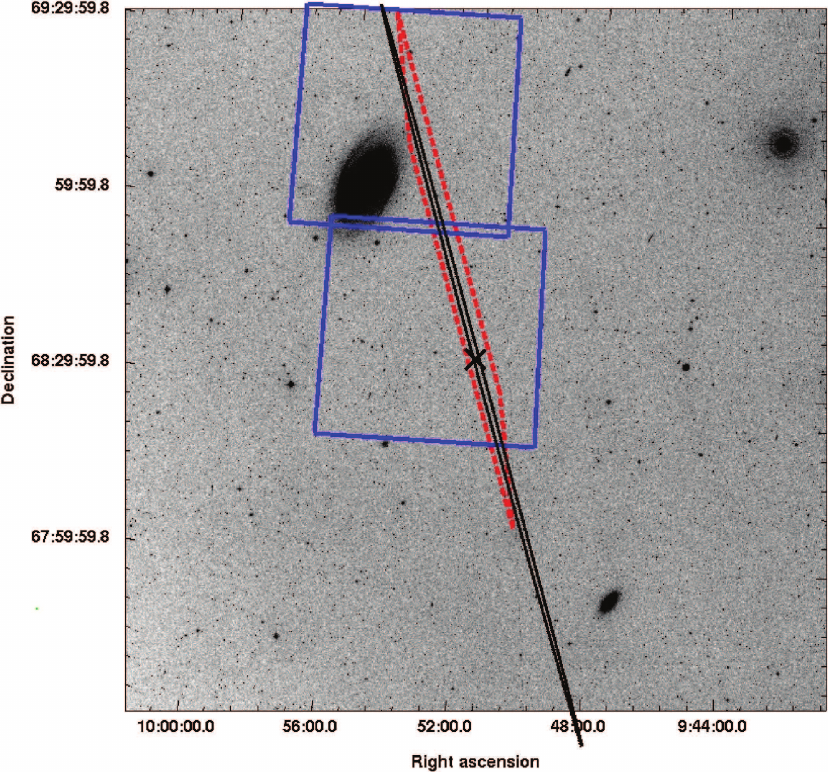}
\caption{The central region of the M81 group, showing the original error
trapezium (red dashed line) from the IPN and the refined $3\mbox{-}\sigma$ error ellipse
(solid black).  The blue boxes are the regions studied in the optical.
Figure from \citet{Hurley10} Copyright (c) 2010 RAS.}
\label{fig:m81}
\end{center}
\end{figure}

GRB~051103 was a short-duration, hard-spectrum gamma-ray burst (GRB) which
occurred at 09:25:42 UTC on 3 November 2005 \citep{Hurley10} and was
possibly located in the nearby galaxy M81, at a distance
3.63$\pm$0.14\,Mpc from Earth\,\citep{gcn4197, durrell10}.  A preliminary
quadrilateral error box obtained by the third interplanetary network of
satellites (IPN3) was consistent with a source in the M81 group
\citep{gcn4197}.  The refined $3\mbox{-}\sigma$ error ellipse, shown with a
solid black line in Figure\,\ref{fig:m81}, has an area of 104 square arcminutes,
and excludes the possibility that the GRB's source was the inner disk of
M81\,\citep{Hurley10}.  The location of the progenitor of GRB~051103 is,
however, consistent with the outer disk of M81.

Two other galaxies are noted to lie within the original error box: 
PGC028505 (distance estimated at 80\,Mpc, \citet{gcn4206}) and PGC2719634
(distance unknown).  PGC2719634 lies on the $18\%$ confidence contour of
the refined ellipse and constitutes a plausible host galaxy.  PCG028505, however,
lies on the $0.03\%$ contour and is unlikely to be the host.  Furthermore,
PGC028505 was observed in the R and V bands but no evidence for brightening due
to an underlying transient source was found \citep{gcn4207} and it is not
thought to be a plausible host of GRB~051103 \citep{Hurley10, gcn4206}.
Observations of the original quadrilateral error box in optical and radio
concluded that GRB~051103 was not associated with any typical supernova at $z
\lesssim 0.15$ \citep{ofek-2006-652}.  None of the known supernova remnants in
M81 fall within the refined elliptical error region.

The progenitors of most short duration GRBs are widely thought to be the
coalescence of a neutron star-neutron star (NS-NS) or neutron star-black hole
(NS-BH) binary system (see, for example, \citealt{nakar-2007} and references
therein). With the right combination of binary
masses and spins, the neutron star matter is believed to be tidally disrupted
leading to the formation of a massive torus.  Accretion of matter from this
torus onto the final post-merger object leads to the formation of highly
relativistic outflows along the axis of total angular momentum of the system
\citep[e.g.,][]{Setiawan2004, Shibata2006, Rezzolla2011}.  Internal shocks in
the relativistic jet give rise to the prompt $\gamma$-ray emission observed in
short, hard GRBs.

These binary systems also produce a characteristic gravitational-wave (GW) signal in the
last few seconds before coalescence that is detectable by the current generation of 
interferometric GW detectors, such as those of the Laser Interferometer
Gravitational-wave Observatory \citep[LIGO,][]{abbottnim04,abbott-2007}, to
$O(10)$\,Mpc.  Thus, if M81 was indeed the host of a binary merger progenitor of GRB~051103,
LIGO should have detected a GW signal associated with the event.
A similar hypothesis was tested for GRB 070201, whose error box
overlapped the spiral arms of M31 (which is 770~kpc from Earth).  LIGO was able
to exclude a compact binary progenitor in M31 at $>99\%$ confidence
\citep{grb070201_07} and placed a lower limit of 3.5\,Mpc on its distance at
$90\%$ confidence.

Up to $15\%$ of short GRBs might be giant flares from 
soft gamma repeaters (SGRs) in the local universe~\citep{Tanvir2005,Chapman2009}.
SGRs are believed to be magnetars; neutron stars with extremely large magnetic
fields ($B\sim10^{15}$\,G).  
However, only a few percent of short GRBs are thought to share 
the SGR-like properties exhibited by GRB 051103 \citep{frederiks-2007}.
For example, the light curve exhibits the steep rise ($\sim 4$\,ms) and 
decaying tail observed in the initial pulses of SGR giant
flares \citep{frederiks-2007, ofek-2006-652, Hurley10}.
At the distance of M81 the characteristic late-time weaker, oscillatory phase
expected of a SGR giant flare, which follows the rotation of the underlying
neutron star, would not be detectable \citep{Hurley10}.  Second, the
spectrum of GRB 051103 shows the hard-to-soft evolution characteristic of
SGR giant flares \citep{frederiks-2007}.  Also, if we assume the source was in
M81, the isotropic electromagnetic energy release is approximately $3.6 \times
10^{46}$\,erg\,\citep{gcn4197}, consistent with the energy release
($\sim\sci{4}{46}$\,erg) of the SGR 1806\textminus20 giant
flare\,\citep{hurley05}.  We note that a number of UV-bright regions contained
within the elliptical error box indicate star-forming regions in the outer disk
of M81 which may host magnetars \citep{ofek-2006-652,Hurley10}.  If
confirmed, the identification of an SGR in M81 would be the second and most
distant extra-galactic SGR flare observed to date.

Several searches for GWs associated with magnetar events have already been
performed \citep{LIGOSGR1806-20,LIGOS5SGR,LIGOSGR1900+14,s5y2A5sgr}.  No
evidence of a GW signal was found in these searches, including the 2004 giant
flare from the Galactic magnetar SGR 1806\textminus20, which is a factor of $\sim$300
closer to Earth than M81\,\citep{LIGOS5SGR}.  A detectable GW signal from a
magnetar giant flare in M81 would therefore probably require $>10^5$ more energy
in the GW emission over SGR 1806\textminus20.  

At the time of the GRB, the LIGO detectors were in final preparations for their
fifth science run, S5,  which began the following day.  For this reason, the data
from around the time of GRB~051103 has not been included in previous searches
associated with GRBs or SGRs in S5 data.  Nonetheless, data taken by the LIGO
2\,km detector in Hanford, WA (H2) and the LIGO 4\,km detector in Livingston,
LA (L1) is available.  Motivated by interest from the astronomical community
and the potential for a GW detection, we have performed a search using the
established data analysis pipelines from the S5 searches.

The data was calibrated as described in \citet{Abadie2010223}.  
The validity of the calibration was established by comparing records of the 
detector configuration at the GRB epoch to those near the start of the science 
run, and estimates of the calibration uncertainty are accounted for 
in the GW searches.
Data quality studies and techniques for vetoing problematic segments were
similar to those used during S5 \citep{S5firstyear}.  These detector
characterization studies have established that the data is of science quality
and equivalent to that shortly after the official start of S5.

In this paper we report on the LIGO search for GWs associated with GRB~051103, and
the resulting implications for the origin of this GRB\@.  Three independent analysis
packages, designed for different purposes, were used.  In \S~\ref{sec:cbc} we
describe the method and results of searching for theoretically predicted
gravitational waveforms emitted during compact binary mergers.  In
\S~\ref{sec:burst}, we describe the results of two searches using analyses
which are designed to be sensitive to unmodelled short-duration ($\lesssim
1$\,s) bursts of GWs.  The first is an analysis designed to search for GW bursts
from magnetar flares.  The second performs a search for generic GW
bursts from GRBs in the sensitive band of the LIGO instruments.  Finally, we
summarise our findings in \S~\ref{sec:conclusion}.

\section{Search for GWs from a compact binary progenitor}
\label{sec:cbc}

\subsection{Search Method}
\label{sec:method}

The method used to search for the GW signal from binary coalescence is
identical to that reported in \citet{CBCGRB}: matched filtering is used to
correlate theoretically motivated template waveforms with the data streams from
the detectors.  

The GW signal from binary coalescence is expected to precede the prompt
$\gamma$-ray emission by no more than a few seconds.   We therefore search for
GW signals whose end time lies in an {\it on-source} window of $[-5,+1)$\,s
around the reported GRB time.  The significance of candidate GW signals is
estimated from the background distribution of 324 {\it off-source} trials, each
$6$\,s long (the number of which is dictated by the quality of the data around 
the time of the GRB).  

The form of the GW signal from compact binary coalescence depends on the 
masses ($m_\mathrm{NS}, m_\mathrm{comp}$) and spins of the neutron star 
and its companion (either a neutron star or a black hole), as well as the 
spatial location relative to the detectors, the inclination angle $\iota$ 
between the orbital axis and the line of sight, and
the polarization angle specifying the orientation of the orbital axis.
The data from each detector is filtered through a discrete bank of template
waveforms designed such that the maximum loss of signal-to-noise ratio (SNR)
due to discretization effects for a binary with negligible spins is $3\%$.  
Although the template bank used
ignores spin, we later evaluate our sensitivity to spinning systems and verify
that such systems are still detectable.  It is assumed that at least one member
of the binary is a neutron star with mass $1\,M_{\odot} \leq m_{\rm NS}\leq
3\,M_{\odot}$.  For the companion object, we test masses in the range
$1\,M_{\odot} \leq m_{\mathrm{comp}} \leq 25\,M_{\odot}$ to allow the
possiblity of a either a neutron star-neutron star or neutron star-black hole
merger.  We note that black hole masses greater than $25\,M_{\odot}$ seem
likely to ``swallow'' the neutron star whole, without tidally disrupting the
neutron star and forming a sufficiently massive accretion disk to power a GRB
jet \citep{Belczynski08}.

If the matched filter SNR exceeds a threshold, 
the template masses and the time
of the maximum SNR are recorded.  These \emph{triggers} between detectors are
then tested for coincidence in their time and mass
parameters \citep{Robinson:2008}. This significantly reduces the number of
background triggers that arise from matched filtering in each detector
independently.  
Further background suppression is achieved by applying signal consistency 
tests, specifically a $\chi^2$ test \citep{Allen:2004} and the $r^2$ 
veto \citep{Rodriguez:2007}.
The SNR and $\chi^2$ from a single detector are combined into an
effective SNR \citep{LIGOS3S4all}, which is then summed in quadrature 
across detectors to form the combined effective SNR which is used as 
the ranking statistic.

The distribution of effective SNRs can vary significantly across the range of
masses being searched, with shorter, higher mass templates more susceptible to
non-stationary background noise.  Consequently, we split up the search space by
mass and re-rank triggers in each mass bin by their likelihood of having arisen due to a
gravitational wave signal.  This is defined as the efficiency with which we
detect plausible gravitational wave signals divided by the false-alarm
probability, for a given combined effective SNR\@. The false-alarm probability
is the probability of obtaining a  candidate louder than that observed in the
on-source trial in the same region of mass space from noise alone; it is
measured using the off-source trials.  The detection efficiency is computed by
adding simulated gravitational wave signals to the data from off-source trials
and counting the fraction which are recovered by the detection pipeline.  

\subsection{Search Results} 

The matched-filter search found no evidence for a GW signal produced by compact
binary coalescence at the time of GRB 051103.  The most significant candidate
event in the on-source region around the time of the GRB had a false alarm
probability of $\onsourceFAP\%$.  That is, there was a $\onsourceFAP\%$ chance
of observing a candidate this loud or louder in any given off-source trial due
to an accidental coincident noise fluctuation in each detector.

The null-detection result allows us to compute the frequentist confidence with
which we may exclude binary coalescence in M81 as the progenitor for this GRB\@.
We used the approach of \citet{Feldman:1997qc} to compute regions in distance
where GW events would, with a given confidence, have produced results
inconsistent with our observations.  The Feldman-Cousins confidence regions are
computed by analyzing a family of simulated gravitational wave signals, with a
choice of priors for the intrinsic parameters motivated by astrophysical
observations.  Results are quoted explicitly in terms of either a fiducial
NS-NS or NS-BH merger.

In the case of NS-NS mergers, masses are drawn from a Gaussian distribution with
mean $1.4\,M_{\odot}$, standard deviation $0.2\,M_{\odot}$ and truncated at
$[1.0,3.0]\,M_{\odot}$.  The dimensionless spins, $a=Jc/GM^2$, where $J$ is the
spin angular momentum and $M$ is the mass, are uniformly distributed within
$[0.0,0.4]$.  The upper bound is chosen to be compatible with the spin of the
fastest observed millisecond pulsar \citep{Hessels2006}.
Our fiducial NS-BH systems have black hole masses drawn from Gaussian with mean
$10.0\,M_{\odot}$, standard deviation $6.0\,M_{\odot}$ and truncated at
$[2.0,25.0]\,M_{\odot}$. To reflect the greater uncertainty arising from a lack
of observed NS-BH systems, the neutron star mass is drawn from a Gaussian
distribution with mean $1.4\,M_{\odot}$ and standard deviation $0.4\,M_{\odot}$.
Black hole spins are distributed uniformly within $[0.0,0.98)$.  Additionally,
population synthesis studies of NS-BH mergers appear to indicate that the tilt
angle (the angle between the BH spin direction and the NS orbital axis) must be
$<45^{\circ}$ in most systems to allow for tidal disruption of the NS and
formation of a sufficiently massive torus able to power the gamma-ray
burst \citep{Belczynski08}.  Recent numerical simulations of
NS-BH mergers lend support to this restriction and find that the tilt angle is
likely $<60^{\circ}$ \citep{Rantsiou2010,Foucart2011}.  We restrict the tilt
angle to be $<60^{\circ}$.

The outflows from the accretion jets in a GRB are directed along the rotational
axis of the final object.  Relativistic beaming and collimation due to the
ambient medium confines the jet to a semi-angle $\theta_{\rm jet}$.  The
observation of prompt $\gamma$-ray emission is, therefore, indicative that the
inclination of the total angular momentum with respect to the line of sight
lies within the jet cone.   Estimates of $\theta_{\rm jet}$ are based on jet
breaks observed in X-ray afterglows and vary across GRBs.  Indeed, many GRBs do
not even exhibit a jet break.  However, studies of observed jet breaks in Swift
GRB X-ray afterglows find a mean (median) value of $\theta_{\rm jet} =
5.4^{\circ} (6.4^{\circ})$, with a tail extending almost to
$25^{\circ}$ \citep{Racusin2009}.  In at least one case where no jet
break is observed, the inferred lower limit is $25^{\circ}$ and could be as
high as $79^{\circ}$ \citep{Grupe2006}.  In order to probe the range of
predicted jet opening angles, we perform separate sets of simulations where the
inclination of the total angular momentum is restricted to jet semi-angles of
$10^{\circ},20^{\circ},\dots,60^{\circ}$ and $90^{\circ}$, allowing an estimate
of exclusion confidence as a function of jet semi-angle.

Systematic errors are treated identically to those in \citet{CBCGRB}: amplitude
calibration uncertainty and Monte-Carlo counting statistics from injections are
the dominant errors.  Amplitude calibration uncertainty is accounted for by
multiplying exclusion distances by $1.28 \times (1 + \delta_{\rm cal})$, where
$\delta_{\rm cal}$ is the overall fractional uncertainty in amplitude
calibration, estimated at $25\%$.  This is significantly larger than typical
science run calibration uncertainties (see e.g., \citet{Abadie2010223}) as fewer
calibration measurements were available from this pre-science run time. 
The factor of 1.28 corresponds to a $90\%$ pessimistic fluctuation, assuming 
Gaussianity.  We incorporate Monte-Carlo uncertainties from the computationally
limited number of simulations by stretching the Feldman-Cousins
confidence regions to cover a probability interval
$\mathrm{CL}+1.28\sqrt{\mathrm{CL}(1-\mathrm{CL})/n}$, where $\mathrm{CL}$ is
the desired confidence limit and $n$ is the number of simulations used in
constructing the interval.

\begin{figure}
\includegraphics{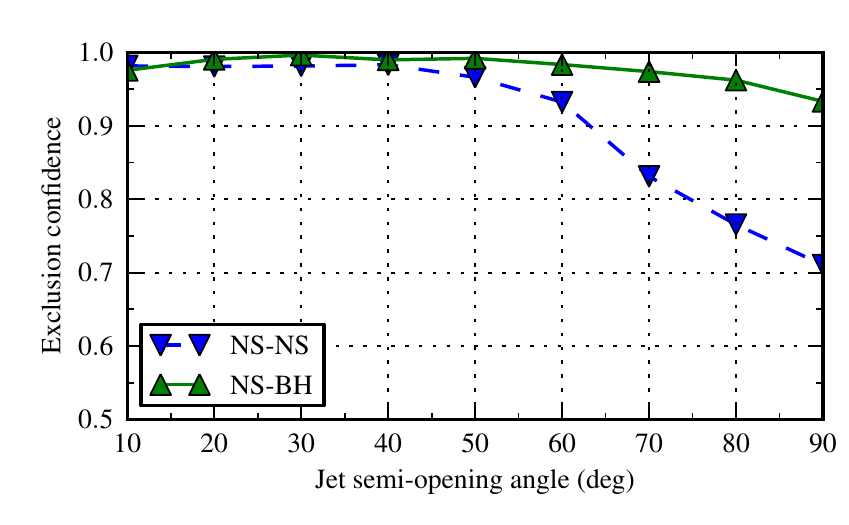}
\caption{Exclusion confidences for the two classes of
compact binary coalescences considered in the matched-filter analysis as a
function of jet semi-opening angle and assuming a distance of $3.63$\,Mpc to
GRB\,051103.  The estimate is based on simulations where neutron
star masses are Gaussian distributed with mean $1.4\,M_{\odot}$ and standard
deviation $0.2\,M_{\odot}$.  Black hole masses are also Gaussian distributed with 
mean $10.0\,M_{\odot}$ and standard deviation $6.0\,M_{\odot}$.  The
reduced confidence below $30^{\circ}$ is purely due to numerical
corrections for limited simulation size.} \label{fig:cbc exclusions}
\end{figure}

Figure~\ref{fig:cbc exclusions} shows exclusion confidence for NS-NS and NS-BH
mergers as a function of jet semi-angle $\theta_{\rm jet}$, assuming a distance
to M81 of $3.63$\,Mpc.  If we assume isotropic $\gamma$-ray (i.e., unbeamed)
emission from GRB~051103 the possibility of NS-NS coalescence in M81 as its
progenitor is excluded with $\BNSExc\%$ confidence.  Taking a fiducial jet
semi-angle of $\theta_{\rm jet}=30^{\circ}$, exclusion confidence rises to
$\BNSExcFaceOn\%$.  NS-BH mergers with isotropic emission are excluded at
$\NSBHExc\%$ confidence, rising to $\NSBHExcFaceOn\%$ for $\theta_{\rm
jet}=30^{\circ}$.  

\begin{figure}
\includegraphics{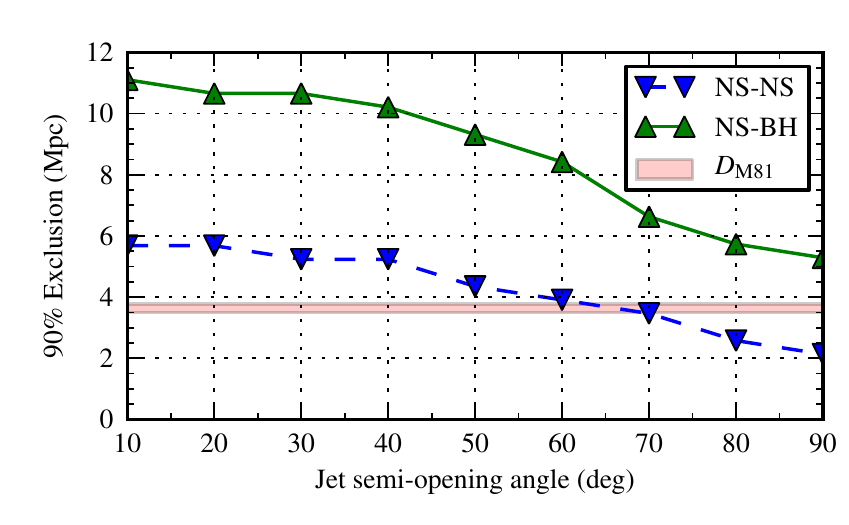}
\caption{$90\%$-confidence exclusion distance as a function of jet semi-angle
for binary coalescences, given LIGO observations at the time of GRB~051103.
\label{fig:distance exclusions}} \end{figure}

To address how far we can exclude binary coalescences if GRB~051103 was
\emph{not} in M81, figure~\ref{fig:distance exclusions} shows the distance at
which we reach $90\%$ exclusion confidence as a function of jet semi-angle.
Assuming unbeamed emission, NS-NS mergers are excluded with $90\%$ confidence
out to a distance of $\BNSDninety$\,Mpc, rising to $\BNSDninetyFaceOn$\,Mpc for
$\theta_{\rm jet}=30^{\circ}$.  The corresponding distances for NS-BH
coalescences are $\NSBHDninety$\,Mpc and $\NSBHDninetyFaceOn$\,Mpc,
respectively.  

The increase in exclusion confidence for smaller jet angles is
due to the fact that the average amplitude of the GW signal from compact binary
coalescence is smaller for systems whose orbital plane is viewed
`edge-on' (where the detector receives the flux from just one GW polarization) 
than for systems viewed `face-on' (where the detector receives the flux from both
GW polarizations); small jet angles imply a system closer to face-on.

\section{Search for a GW burst}
\label{sec:burst}
 
\subsection{Search Methods}\label{sec:burst_method}

We perform two searches for a GW burst associated with GRB~051103.  
As discussed previously, there is evidence that a fraction of 
short GRBs are caused by nearby magnetar flares, so we
perform a search tailored to the expected GW signal arising from such a
flare.  Additionally, we perform a search for a generic GW burst in the
time around the GRB\@.

The \flare{} pipeline \citep{kalmus07, kalmus08} targets neutron star
fundamental mode ($f$-mode) ringdowns as well as unmodeled short-duration GW
signals. It has been used previously to search for GWs associated with Galactic
magnetar bursts including the December 2004 giant flare from SGR 1806\textminus20
\citep{LIGOS5SGR,LIGOSGR1900+14,s5y2A5sgr}.  As in the previous magnetar
searches, we use an on-source region of $[-2,+2]$\,s about the GRB~051103
trigger, and an off-source region of 1000\,s on either side of the on-source
region to estimate the significance of on-source events.

{\flare} produces a time-frequency pixel map from the conditioned and
calibrated detector data streams in the Fourier basis, groups pixels
using density-based clustering, and sums over the group to produce
events.  The data from each of the two detectors is combined by including
detector noise floor measurements and antenna responses to the source
sky location as weighting factors in the detection statistic.  We divide
the search into three frequency bands:  1--3\,kHz where $f$-modes are
predicted to ring; and 100--200\,Hz and 100--1000\,Hz where the detectors are
most sensitive.  In the $f$-mode band we use a Fourier transform length of
250\,ms, which we find to be optimal for $f$-mode signals expected to decay
exponentially with a timescale $\tau$ in the 100--300\,ms range \citep{benhar04}.

The \xp{} analysis package \citep{Sutton:2009gi} searches for generic GW 
bursts in data from arbitrary networks of detectors.  \xp{} was previously 
used in the search for GW bursts associated with GRBs in LIGO science 
run 5 and Virgo science run 1, in 2005--2007 \citep{S5BurstGRB}.  Since the 
analysis is not based on a specific GW emission model, we keep the search 
parameters broad to allow for a generic GW burst.  In particular, we define 
our on-source region as the interval $[-120,+60]$\,s around the GRB trigger; 
this conservative window is large enough to accommodate the time delay 
between a GW signal and the onset of the gamma-ray signal in most GRB 
progenitor models.  We use $1.5$~hours of data on either side of the 
on-source region as the off-source region for background characterization.
The frequency band of the \xp{} search is 64--1792\,Hz.

\xp{} combines the data streams from each detector with weighting determined by the 
sensitivity of each detector as a function of frequency and sky position.
This yields time-frequency maps of the signal energy in each pixel.
Candidate GW events are identified as the loudest 1\% of pixels in the map. 
Each is assigned a significance based on its energy and time-frequency volume, 
using a $\chi^{2}$ distribution with two degrees of freedom.  
These candidates are then refined by comparing the degree of correlation 
between the H2 and L1 data streams, rejecting low-correlation events as 
background.  
Surviving events are ranked by their significance, then each is 
assigned a false-alarm probability by comparison to events from the 
off-source region.

\subsection{Search Results}

Neither the {\flare} magnetar search nor the \xp{} analysis yield evidence
for a plausible gravitational-wave burst signal associated with GRB
051103.  Consequently, we place model-dependent 90\% confidence level upper
limits on the isotropic energy emission in gravitational waves $\egw$
associated with this GRB\@.

The limits from the {\flare} analysis were obtained for twelve types of
simulated GW signals: eight $f$-mode ringdowns with circular and linear
polarizations and decay times $200$\,ms; and four band- and time-limited white
noise bursts with durations of 11\,ms or 100\,ms and spanning the 100--200\,Hz
and 100--1000\,Hz bands.  Uncertainties from detector calibrations and Monte
Carlo statistics were folded into upper limit estimates as described in
\citet{s5y2A5sgr}. The best (lowest) upper limit on $\egw$ was
$\sci{2.0}{51}$\,erg, for 100--200\,Hz white noise bursts lasting 100\,ms.  The
lowest $f$-mode upper limit was $\sci{1.6}{54}$\,erg, for circularly polarized
ringdowns at 1090\,Hz.  These are the lowest frequency signals of each
morphology; limits obtained for the other ten simulated signals scale with the noise
floor of the LIGO detectors as expected and the simulations provide a check of
the robustness of the analysis to the large uncertainty in the frequency of a
putative GW signal \citep[for more details on the simulations used see][]{s5y2A5sgr}.

The upper limits produced by the broad \xp{} search are computed in
a similar way, although they make use of circularly polarized 
sine-Gaussian waveforms at 150\,Hz and 1000\,Hz 
\citep[for details, including handling of uncertainties, see][]{S5BurstGRB}.  
We find upper limits on $\egw$ of 
$\sci{1.2}{52}$\,erg at 150\,Hz and $\sci{6.0}{54}$\,erg at 1000\,Hz.
We can convert these results to lower limits on the distance to 
GRB~051103 as a function of $\egw$, giving 
4.4\,Mpc $(\egw/0.01M_{\odot}c^2)^{1/2}$ at 150\,Hz and 
0.20\,Mpc $(\egw/0.01M_{\odot}c^2)^{1/2}$ at 1000\,Hz.

Even near the frequency of LIGO's best sensitivity our energy upper limits are
several orders of magnitude larger than the maximum energy available for
emission by SGRs in gravitational waves $\sim 10^{46}$--$10^{49}$\,erg
\citep{pacheco98,2001MNRAS.327..639I,Owen:2005fn,horvath05,Corsi2011}.  Indeed,
the energy actually emitted as gravitational waves may be much less than
this~\citep{Kashiyama2011, Levin2011}.  We are
therefore unable to inform the hypothesis of an SGR progenitor for GRB 051103.

\section{Conclusion}
\label{sec:conclusion}

We analyzed data from the LIGO L1 and H2 GW detectors in a frequency band
spanning 40\,Hz to 3\,kHz, looking for a GW signal associated with the
short-hard GRB~051103.  Three data analysis pipelines were deployed, two of
which are designed to search for unmodelled, short-duration ($\lesssim 1$\,s)
burst-like GW signals and one which performs a matched-filter analysis using
templates based on the GW signal expected from compact binary mergers.  
No evidence was found for a GW signal associated with this GRB\@.

The sensitivity of the matched-filter search allows us to confidently 
exclude the hypothesis that the progenitor system was a compact binary 
merger progenitor in M81.  Specifically, assuming an
outflow jet semi-angle $\theta_{\rm jet}=30^{\circ}$, we exclude a NS-NS merger
in M81 at $\BNSExcFaceOn\%$ confidence.  NS-BH mergers with similarly beamed
emission are excluded at $\NSBHExcFaceOn\%$ confidence.  Relaxing the
assumption of beaming such that we include systems whose orbital plane is
oriented edge-on to our line-of-sight, the confidences for NS-NS and NS-BH
mergers fall to $\BNSExc\%$ and $\NSBHExc\%$, respectively.  As a measure of
the distance to which we are sensitive to such events, the $90\%$-confidence
exclusion distances for NS-NS and NS-BH systems with beaming are
$\BNSDninetyFaceOn$\,Mpc and $\NSBHDninetyFaceOn$\,Mpc, respectively.  Assuming
no beaming, these distances drop to $\BNSDninety$\,Mpc and $\NSBHDninety$\,Mpc.

The null result of the searches for an unmodelled burst of GWs allows 
us to set upper limits on the GW energy emission of GRB~051103.  
These limits are in the range $10^{51}$--$10^{55}$\,erg, depending 
primarily on the assumed GW frequency.  These limits are several orders of
magnitude greater than the maximum observed electromagnetic emission from SGRs, $\sim
10^{46}$\,erg, and the highest predictions of the available resevoir of energy
available for gravitational wave emission, so we are not able to constrain 
the hypothesis of an SGR progenitor for GRB~051103.

We conclude then, that it is highly unlikely that the progenitor for
GRB~051103 was a compact binary merger in M81.  If the event indeed occurred
in M81, it seems likely on the basis of LIGO observations that this was indeed
the most distant SGR giant flare observed to date.

\acknowledgments
The authors thank A.~Rowlinson and N.~Tanvir for bringing GRB~051103 to our attention.
The authors gratefully acknowledge the support of the United States National
Science Foundation for the construction and operation of the LIGO Laboratory and
the Science and Technology Facilities Council of the United Kingdom, the
Max-Planck-Society, and the State of Niedersachsen/Germany for support of the
construction and operation of the GEO600 detector. The authors also gratefully
acknowledge the support of the research by these agencies and by the Australian
Research Council, the International Science Linkages program of the Commonwealth
of Australia, the Council of Scientific and Industrial Research of India, the
Istituto Nazionale di Fisica Nucleare of Italy, the French Centre National
de la Recherche Scientifique, the Spanish Ministerio de
Educaci\'on y Ciencia, the Conselleria d'Economia, Hisenda i Innovaci\'o of the
Govern de les Illes Balears, the Royal Society, the Scottish Funding Council,
the Scottish Universities Physics Alliance, The National Aeronautics and Space
Administration, the Carnegie Trust, the Leverhulme Trust, the David and Lucile
Packard Foundation, the Research Corporation, and the Alfred P. Sloan
Foundation.
This document has been assigned LIGO Laboratory document
number {LIGO-P100}{00097}-v\dccversion.

\bibliography{GRB051103_paper}

\end{document}